\documentclass{iopart}

\usepackage{graphicx}

\newcommand{\eV}{\,\mathrm{eV}}

\newcommand{\Mpc}{\,\mathrm{Mpc}}
\newcommand{\ud}{\mathrm{d}}

\begin{document}
\title{A closer look at the spectrum and small scale anisotropies of
UHECRs}

\author{Daniel De Marco\dag, Pasquale Blasi\ddag\ and Angela V Olinto\P}

\address{\dag\ Bartol Research Institute, University of Delaware
Newark, DE 19716, U.S.A.}

\address{\ddag\ INAF/Osservatorio Astrofisico di Arcetri,
Largo E. Fermi, 5 - 50125 Firenze, ITALY}

\address{\P\ Department of Astronomy \& Astrophysics,
Kavli Institute for Cosmological Physics,
5640 S. Ellis Ave. Chicago, IL 60637, U.S.A.}

\eads{\mailto{ddm@bartol.udel.edu}, \mailto{blasi@arcetri.astro.it},
\mailto{olinto@oddjob.uchicago.edu}}

\begin{abstract}
We present results of numerical simulations of the propagation of 
ultra high energy cosmic rays (UHECRs) over cosmological distances,
aimed at quantifying the statistical  significance of the highest energy
data on the spectrum and small scale anisotropies as detected by the
AGASA experiment. We assess the significance of the lack of a GZK
feature and  its compatibility with the reported small scale
anisotropies. 

Assuming that UHECRs are protons from extragalactic sources, we find
that the small scale anisotropies are incompatible with the reported
spectrum at a probability level of $2 \times 10^{-5}$. Our analysis of
the AGASA results shows the power of combining spectrum and small scale
anisotropy data in future high statistics experiments, such as Auger.

\end{abstract}

\maketitle

\section{Introduction}

The problem of the origin of ultra-high energy cosmic rays (UHECRs) is a
collection of several questions that will only be solved by high
statistics observations around $\sim 10^{20}$ eV. The first question is
whether there is a Greisen-Zatsepin-Kuzmin \cite{GZK66} (GZK) feature in
the spectrum of UHECRs due to photopion interactions  with the cosmic
microwave background (CMB). Photopion production is expected to generate
a flux suppression around $\sim 10^{20}$ eV possibly followed by a
recovery. The statistical significance achieved by current experiments
is not sufficient to clarify whether the feature is present and what is
its exact shape (see, e.g., \cite{DBO03}), but future data from the
Pierre Auger Observatory \cite{PAO} should resolve this issue. The
second part of the problem is related to the identification of a class
or several classes of sources that can accelerate particles to energies
in excess of  $10^{20}$ eV. The search for sources of UHECRs will likely
be a long process requiring the identification of anisotropies in the
arrival direction of the highest energy cosmic rays as well as
multiparticle observations of high energy sources (see, e.g.,
\cite{FBD05,ste}). Large scale anisotropies may indicate the generic
host of sources while small scale anisotropies can eventually identify
single sources directly. As small scale anisotropies start to become
apparent with small number of events per source (e.g., doublets and
triples),  an estimate of the  number density of the sources becomes
feasible (see, e.g. \cite{BD04,others}). For a clear source
identification with single source spectral measuments, a future
generation of experiments will be necessary.

If astrophysical sources are responsible for the acceleration of UHECRs,
the mean density of sources can be determined by analyzing small scale
anisotropies (SSA)  in the arrival directions that should become
apparent as large data sets of the highest energy events become
available. So far the only experiment to report SSA is AGASA
\cite{agasa_ssa} and estimates for the number density of sources range
from  $\sim 10^{-6}\rm Mpc^{-3}$  to  $\sim 10^{-4}\rm Mpc^{-3}$ with a
large uncertainty \cite{BD04} (see \cite{others} for other estimates).
In Ref. \cite{BD04},  a full numerical simulation of the propagation was
performed accounting for the statistical  errors in the energy
determination and the AGASA acceptance with the conclusion that the
AGASA small scale anisotropies is consistent with a density of sources
$\sim 10^{-5}\rm Mpc^{-3}$.  This source density corresponds to a
luminosity per source of $\sim 10^{42} \rm erg~s^{-1}$ at energies
$E\geq 10^{19}$ eV and a  best fit injection spectrum in the form of a
power law has a spectral index $\gamma=2.6$ if the source luminosity has
no significant evolution with redshift. For the case of strong source
evolution, in \cite{BD04} it was assumed that the luminosity varies as a
function of redshift as $L(z)\propto (1+z)^4$ and the best fit slope at
injection is $\gamma=2.4$. These fits are dominated by the energy
regions with the largest number of  events, i.e., at $E\sim 10^{19}$ eV,
such that a dominant contribution from galactic sources at these
energies may significantly affect these results. The issue of the energy
region where the transition from a galactic to extragalactic origin
takes place is currently a subject of much debate. While the traditional
view is that the ankle is indeed determined by such a transition, some
recent work \cite{bere1,bere2} have proposed an alternative picture, in
which the ankle is the natural result of the pair production process on
cosmological propagation of extragalactic protons. The key issue in this
class of models is the chemical composition, since a transition from
heavy to light elements is predicted \cite{bere1,bere2}. A small
contamination of the extragalactic component with iron nuclei would
destroy the pair production dip and invalidate this model
\cite{bere2,allard}. If the ankle is due to this mechanism, the data at
energies above and around the ankle are well fit by an injection
spectrum in the form of a power law with slope $2.6-2.7$ and a low
energy cutoff that can be the result of the appearance of a magnetic
horizon \cite{tshorizon,alo,L04}. 

The results on small scale anisotropies, obtained for cosmic rays at
energies above $4\times 10^{19}$ eV, are not significantly affected by a
possible galactic contamination at lower energies. Moreover at such high
energies the effect of the galactic magnetic field is expected to not
affect the propagation in a significant way. 

The role of the intergalactic magnetic field is less constrained as
different simulations give different estimates for the magnitude and
spatial structure of these fields (see, e.g., \cite{dolag04,sigl}). A
homogeneous magnetic field spread through the whole universe would leave
the spectrum of diffuse cosmic rays unchanged with respect to the case
of the absence of magnetic field \cite{aloprop}. Here, we assume that
intergalactic magnetic fields can be neglected for particles with
energies above $4\times 10^{19}$ eV. This assumption holds if magnetic
fields in the intergalactic medium are less than $\sim 0.1~\rm nG$ if
the reversal scale is $1$ Mpc and the small scale anisotropies are
evaluated on angles of $\sim 2$ degrees \cite{BD04}. This magnitude
field is compatible with observational bounds \cite{BBO99} and detailed
numerical simulations in \cite{dolag04} (however, see  \cite{sigl} for
different results). These magnetic field calculations are  the result of
the formation of the large scale structure of the universe. Other
authors considered alternative possibilities, for instance, related to
the early stages of evolution of galaxies \cite{lesch}, but in most
cases the results are very model dependent and are affected by further
evolution of magnetic fields during large scale structure formation. It
is likely that UHECR observations will eventually allow us to measure
the strength of the intergalactic magnetic field and get information
about its structure (see, e.g., \cite{L04,Parizot03}).

In the present paper, we address the significance of a detection of the
GZK feature together with the identification of small scale
anisotropies. We combine these two types of observations by AGASA  in
order to check their compatibility. In \S 2, we briefly describe the
Monte Carlo code used to determine the spectra and the SSA. In \S 3, we
discuss simulated spectra of AGASA and HiRes with regards to  the
presence or absence of a GZK feature. In \S 4, we  calculate the
two-point correlation function for different choices of the angular
binning and combine the spectrum and SSA studies. We conclude in \S 5.

\section{The Monte-Carlo code}

The propagation of UHECRs is simulated using a Monte-Carlo code 
described in \cite{DBO03,BD04}. We assume that UHECRs are protons 
injected with a power-law spectrum by extragalactic sources. The 
injection spectrum in taken to be of the form:
\begin{equation}
  F(E) \propto E^{-\gamma} \exp(-E/E_\mathrm{max}),
\end{equation}
where $\gamma$ is the spectral index and $E_\mathrm{max}$ is the maximum
injection energy at the source. We fix  $\gamma=2.6$ and
$E_\mathrm{max}=10^{21.5}\eV$ since these values reproduce well  the
experimental results in the lower energy, high statistics region $\sim
10^{18.5}$ eV to $\sim 10^{20}$ eV \cite{DBO03}. Here, we focus on
events above $4\times10^{19}\eV$, which are generated at $z\ll 1$,
therefore, source evolution is only marginally relevant \cite{BD04}. We
simulate the propagation of protons from the source to the observer by
including the photo-pion production, pair production, and adiabatic
energy losses due to the expansion of the universe. The photo-pion
production is treated as a discrete energy loss process and is simulated
using the event generator SOPHIA \cite{sophia}.

We use a constrained simulation where particles are propagated until the
number of detected events above some energy threshold reproduces the
experimental results. By normalizing the simulated flux by the number of
events above a given energy, where experiments have high statistics, we
can estimate the fluctuations in the number of events above a higher
energy where experimental results are sparse. In this way, we have a
direct handle on the fluctuations that can be expected in the observed
flux due to the stochastic nature of the photo-pion production and to
cosmic variance.

In choosing the initial particles to be propagated, we employ two
different physical scenarios: a) a continuous distribution of sources;
and b) a discrete distribution of point sources with a given density.
For the continuous distribution  a source distance is generated at
random from a uniform distribution in a universe with
$\Omega_\Lambda=0.7$ and $\Omega_\mathrm{m}=0.3$. In an Euclidean
universe, every spherical shell contributes equally to the generated
flux of particles.  The flux from a source scales as $r^{-2}$, where $r$
is the distance between the source and the observer, while the number of
sources between $r$ and $r+\mathrm{d}r$  scale as $r^2$, so that the
probability that a given event has been generated by a source at
distance $r$ is independent from $r$.  In a flat universe with a
cosmological constant this is still true provided the distance $r$ is
taken to be
\begin{equation}\label{eq:r}
   r=c\int^{t_0}_{t_\mathrm{g}}\frac{\ud t}{R(t)} \ ,
\end{equation}
where $t_\mathrm{g}$ is the age of the universe when the event was
generated, $t_0$ is the present age of the universe, and $R(t)$ is the
scale factor of the universe. Once a particle has been generated, the
code propagates it to the detector calculating energy losses and taking
into account the energy and angular resolution of the detector.

For a discrete distribution of point sources, we first generate the
positions of sources with a given spatial density and then emit
particles choosing randomly from the source positions. The source
redshifts are generated with a probability distribution proportional to
\begin{equation}
   \frac{\ud n}{\ud z} \propto r(z)^2 \frac{\ud t}{\ud z}
\end{equation}
where $r(z)$ is the $r$ defined in Eq.~(\ref{eq:r}) and  $\ud t/\ud 
z$ gives the relation between time and redshift for a flat universe with
a cosmological constant (see, e.g.,  the expression given in \cite{bg}).
The source positions on the celestial sphere are chosen uniformly in
right ascension and with a declination distribution proportional to
$\cos\delta$.\footnote{The declination $\delta$ is defined as being
$0^\circ$ on the equatorial plane and $\pm90^\circ$ at the poles, so
$\ud \Omega = \cos\delta \ud \delta \ud \alpha$.} 

Since we neglect the effects of the magnetic fields on the propagation,
we ignore sources outside the experimental field of view and assign to
visible sources a probability proportional to:
\begin{equation}
   r(z)^{-2} \omega(\delta)\,,
\end{equation}
where $r(z)^{-2}$ takes into account the distance dependence of the
solid angle and $\omega(\delta)$ is the relative exposure of the
experiment in the given direction. The positions for particle injection
are chosen  randomly from the assigned source positions and propagation
proceeds as in the previous case.

The last point worth stressing is the importance of the statistical
errors in the energy determination. The statistical errors in the energy
are  accounted for in our simulation by assigning a {\it detection}
energy chosen at random from a gaussian distribution centered at the
arrival energy $E$ and with width $\Delta E/E = 30\%$. As we show in the
next section and in  \cite{DBO03}, the presence of a statistical error
in the energy determination does affect the shape of the spectrum of
UHECRs close to the GZK feature. This point was also made recently  in
\cite{Albuq05} for the case of Log-normal errors in the energy
determination.

\section{Simulated spectra of AGASA and HiRes}

\begin{figure}
   \centering
   \includegraphics[width=0.9\textwidth]{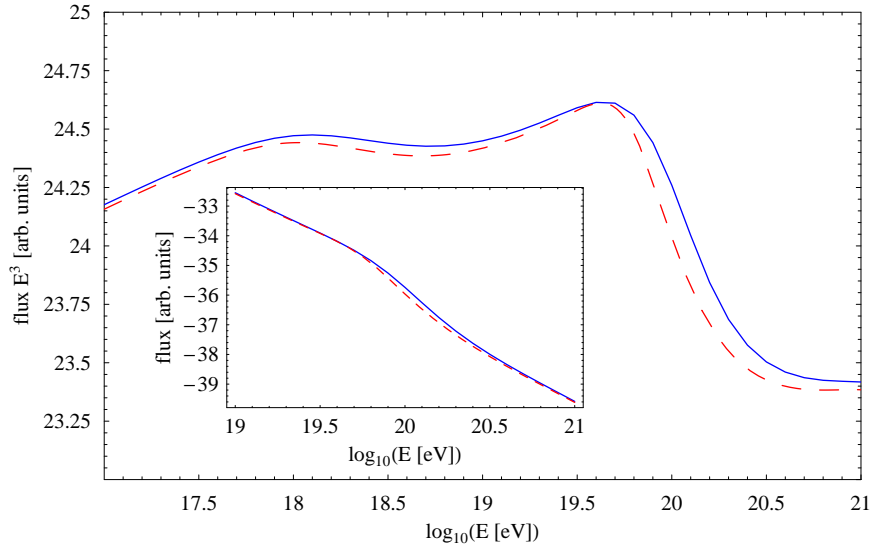}
   \caption{Effect on the spectrum of statistical errors on the energy 
determination. Outer plot: flux multiplied by the third power of the 
energy. Inner plot: flux. Dashed line:  diffuse spectrum with no 
statistical error on the energy determination. Solid line: diffuse 
spectrum with 30\% statistical error on the energy
determination.}\label{fig:staterr} 
\end{figure}

There are several reasons to simulate the expected spectrum of UHECRs,
rather than an analytical treatment, including the ability to introduce
detector-related features which can affect the results. For instance,
the energy dependence and zenith angle dependence of the acceptance of
the experiment can be included as well as statistical and systematic
errors associated with the energy detemination of  each event.  In order
to illustrate the importance of this point, we calculated the spectrum
of UHECRs from a uniform distribution of sources with and without
statistical errors. In Fig. \ref{fig:staterr}, we plot the expected
fluxes (small square inside the main plot) with (solid line) and without
(dashed line) 30\% statistical error in the energy determination. The
main plot illustrates the flux multiplied by the third power of the
energy without statistical error (dashed line), and with a 30\%
statistical error in the energy determination (solid line).

From Fig.~\ref{fig:staterr}, it is clear that the effect of the
statistical error on the flux multiplied by $E^3$ is to smear the shape
of the GZK feature and to move its position toward higher energies. This
effect is due to the steeply falling spectrum and to its slope changes,
so that the probability of misplacing an event from a lower to a higher
energy bin is larger than in the opposite direction. 

\begin{figure}
   \centering
   \includegraphics[width=0.45\textwidth]{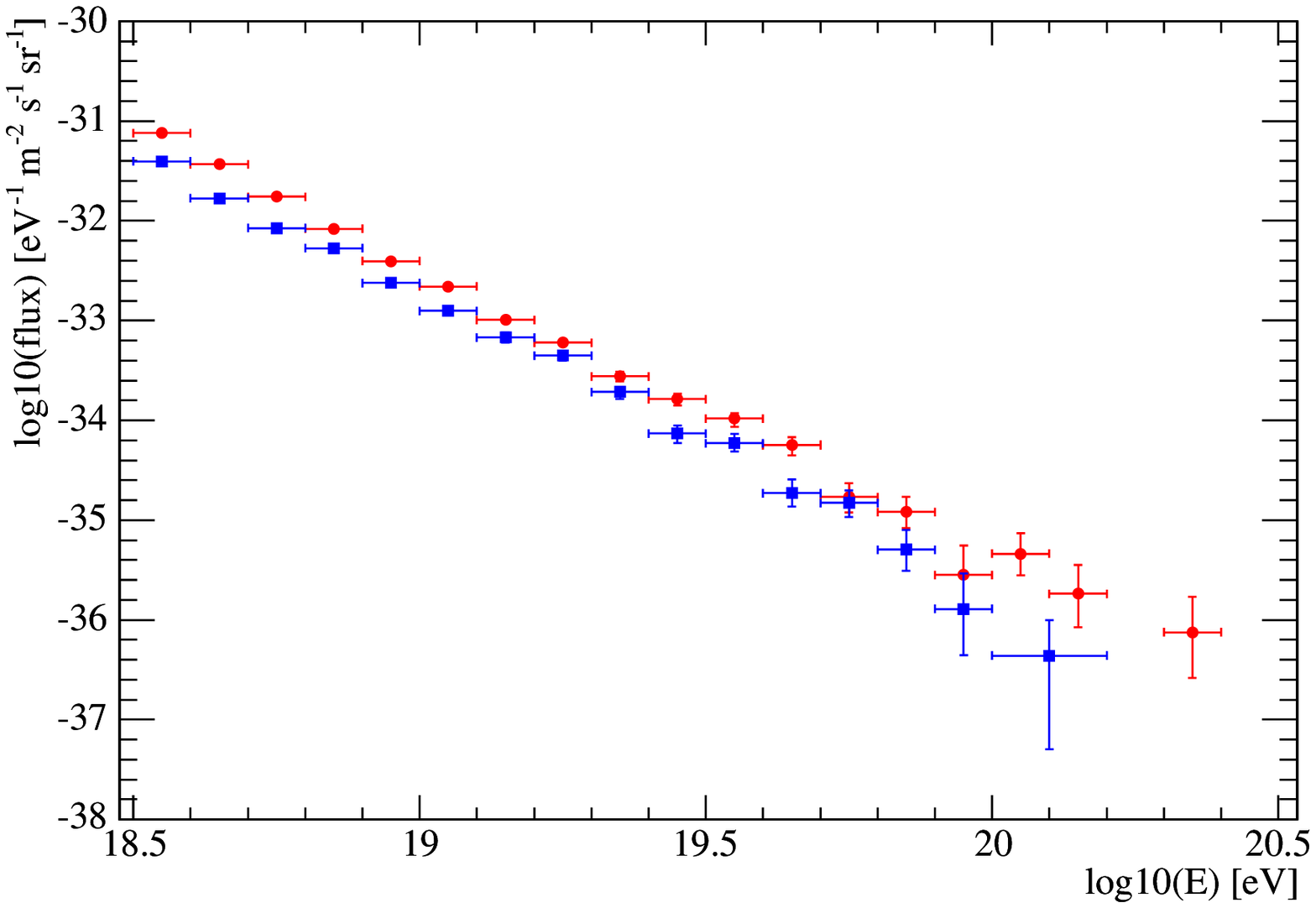}\qquad
   \includegraphics[width=0.45\textwidth]{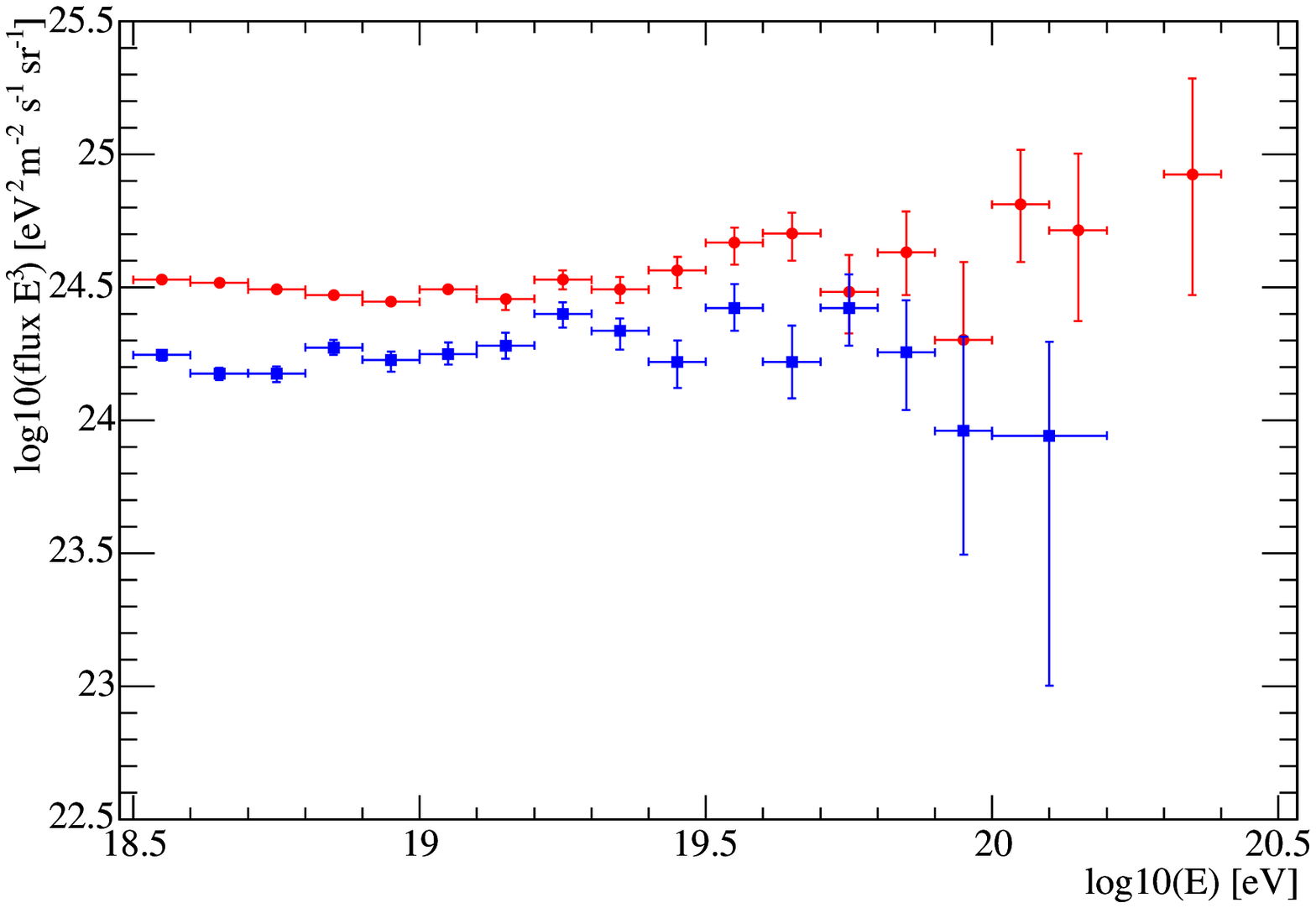}
   \caption{UHECRs spectra as measured by AGASA \cite{agasa_data} 
(circles) and HiResI \cite{hires_data} (squares). Left panel: normal 
spectra. Right panel: spectra multiplied by the third power of the 
energy.}\label{fig:fluxes}
\end{figure}

Results from the two largest exposure  experiments, AGASA and HiRes, are
claimed to be inconsistent. The fluxes from AGASA and HiResI  are
plotted in the left panel of Fig. \ref{fig:fluxes}, while the fluxes
multiplied by $E^3$ are in the right panel. The flux figure suggests
that  despite the different techniques used by the two experiments, they
appear to be in rather good agreement. There seems to be a systematic
shift and some discrepancy at the highest energy bins. Their
discrepancies are made more evident in the right panel, due to the $E^3$
amplification factor. At energies $E>10^{20}$ eV, the HiRes spectrum
shows a hint of a cutoff, while the AGASA spectrum continues unabated.
In Ref.~\cite{DBO03} we showed that a systematic overestimate of the
AGASA energies by 15\% and an underestimate of the HiRes energies by the
same amount would in fact bring the two data sets in much better
agreement in the region of energies below $10^{20}\eV$.  At higher
energies, the very low statistics of both experiments hinders any
statistically significant claim for either detection or the lack of the
GZK feature.

\begin{figure}
   \centering
   \includegraphics[width=0.9\textwidth]{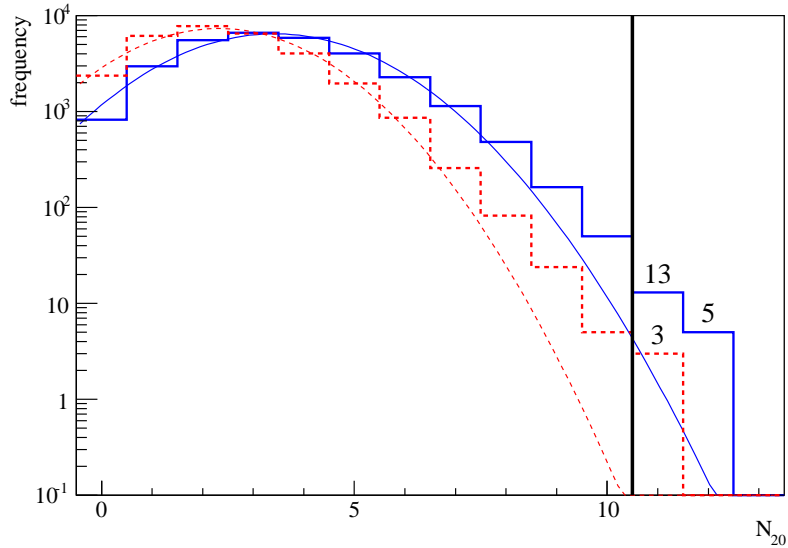}
   \caption{Histogram of the number of detected events with energy above
   $10^{20}\eV$ in 30000 realizations of the AGASA statistics. Dashed
   line: with no statistical error in the energy determination. Solid
   line: with 30\% statistical error in the energy determination. The
   vertical line corresponds to the 11 AGASA events. The numbers above the
   last two bins are the number of events in those bins. The thin lines
   are gaussian fits to the histograms.}\label{fig:n20}
\end{figure}

To clarify the dangers of low statistics, we simulated the AGASA
spectrum normalizing  at $4\times10^{19}\eV$ (instead of $10^{19}\eV$ as
in Ref. \cite{DBO03})  without  the statistical error on the energy
determination and with a 30\% error.  We generated 30000 realizations of
spectra with 72 events above $4\times10^{19}\eV$  like the AGASA data
and recorded the number of events with energies above $10^{20}\eV$ in
each realization. When the experimental error is taken into account, we
obtained $3.5\pm 1.8$ events above $10^{20}\eV$. In the  case of no
errors, the number is $2.5\pm 1.6$. 

Assuming a gaussian distribution with  the above numbers,  we would
incorrectly conclude that the 11 AGASA events above $10^{20}\eV$ are a
$4.2\sigma$ discrepancy, corresponding to a probability of $10^{-5}$. In
Fig.~\ref{fig:n20}, we plot the histogram of the number of events above
$10^{20}\eV$ in each realization. The dashed histogram is the result of
the simulation without statistical errors, whereas the solid one is the
result of adopting a 30\% statistical error. The thin lines are gaussian
fits to the histograms. As it is clear from  Fig.~\ref{fig:n20}, the
distributions are not gaussian such that the naive estimates above of
the discrepancy and of the probability are incorrect. Of the 30000
realizations we simulated the number of spectra presenting 11 or more
events with energy above $10^{20}\eV$ is 18 for the simulations with
statistical errors. This corresponds to a probability of
$6\times10^{-4}$ that in terms of gaussian errors would correspond to a
discrepancy of about $3.2\sigma$. A similar calculation was performed
for the degenerate injection spectrum with  $\gamma=2.4$ and redshift
evolution of the luminosity $L(z)\propto (1+z)^4$. In this case the
probability becomes $5.8\times10^{-4}$,basically identical to the
previous case.

In \cite{DBO03}, we also considered the effect of a possible systematic
error in both the AGASA and HiRes experiments. To match the two spectra,
we assumed that the  AGASA energies had been overestimated by 15\% and
that the HiRes energies were underestimated by the same amount.  In this
case, the number of events above $10^{20}\eV$ in the rescaled AGASA data
set is between 6 and 7, while the number of  events above $10^{19.6}\eV$
is about 47. We simulated 30000 realizations of this statistics above
$10^{19.6}\eV$ and counted the number of events in each realization with
energies above $10^{20}\eV$. In this case, the number of realizations
with 7 or more events above $10^{20}\eV$ turned out to be 179, which
corresponds to a probability of $6 \times 10^{-3}$ or a discrepancy of
$2.5\sigma$ in terms of gaussian errors. Since there are several events
with energy very close to $10^{20}\eV$ we also considered the case of a
number of events above $10^{20}\eV$ equal to 6. Counting the simulated
realizations with 6 or more events above $10^{20}\eV$, we obtained a
probability of 0.024 or about $2\sigma$.

\begin{figure}
   \centering
   \includegraphics[width=0.9\textwidth]{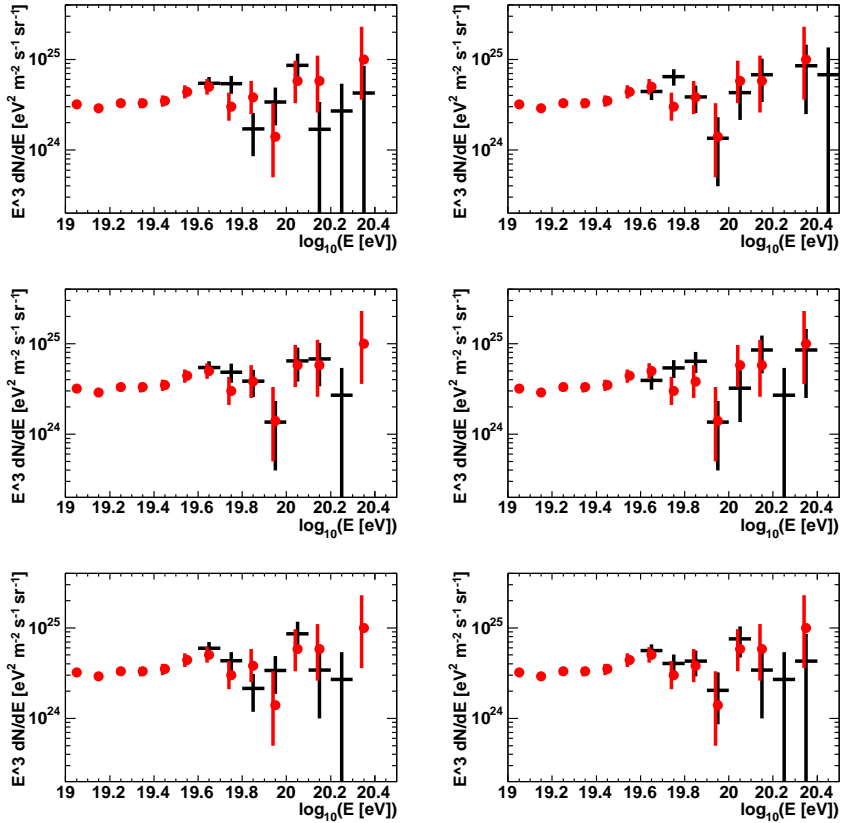}
   \caption{In these  panels we plot 6 of the 18 simulations that have 
11 or more events above $10^{20}\eV$. The black crosses are the 
simulation results. The gray dots with error bars are the AGASA data 
superimposed for comparison. The error bars in the AGASA data are 
slightly shifted left to avoid covering of the black 
crosses.}\label{fig:spectra}
\end{figure}

These statistical considerations are the result of averages over large
samples of realizations of source distributions, therefore one might
wonder whether the spectra of UHECRs for those realizations that have a
large number of events at ultra high energies are characterized by a
pronounced GZK feature or are similar to the AGASA spectrum.  To resolve
this issue we plot, in Fig. \ref{fig:spectra},  the spectra of some of
the realizations that showed 11 or more events above $10^{20}\eV$. These
spectra closely resemble the AGASA spectrum: all of them show no
evidence of a GZK suppression, despite the fact that in the lower energy
region they all fit the  data quite well. This shows that an AGASA-like
spectrum is not that improbable, even if the {\it average} cosmic ray
spectrum can be expected to show a GZK feature.

It is  worth noticing that a few of the AGASA events are just a few
percent above $10^{20}\eV$, so that any small error in the energy
assignment to one or more events would affect the statistical analyses
discussed above. For instance, if just one of the events above
$10^{20}\eV$ had a {\it real} energy below $10^{20}\eV$, the probability
of reproducing the AGASA result would become $2\times10^{-3}$
($2.8\sigma$). On the other hand, if one of the events below
$10^{20}\eV$ had actually an energy above $10^{20}\eV$ then the
probability would become $2\times10^{-4}$ ($3.6\sigma$). So, changing
the number of detected events above $10^{20}\eV$ by just one unit
changes the probabilities of reproducing the AGASA results by relatively
large amounts. These results again  indicate the excessively low
statitics of events, and conclusions concerning the spectrum of UHECRs
at these energies should not be considered as reliable.

This argument applies as well to the case of the HiRes experiment.
Although HiRes data  appears to be consistent with the presence of the
GZK  feature, such a conclusion is also premature. To show this, we
assumed that the actual flux of UHECRs is consistent with the spectrum
measured by AGASA (which is usually considered to be inconsistent with
the GZK feature) and selected events from such spectrum. Then each event
is  either accepted or rejected according to the HiRes aperture
\cite{hiresaperture} (the aperture of HiRes increases with energy while
that of AGASA  is constant at high energies). This procedure is repeated
until the number of accepted events with energy above
$4\times10^{19}\eV$ corresponds to the one measured by HiRes (namely
27). We then count the number of events with energies above $10^{20}\eV$
in each realization and calculate the probability of having one or less
events (namely the HiRes number of detections) in that energy region. We
find that such probability is $\sim 3\%$, corresponding to about
$2\sigma$ in terms of gaussian errors. We also tried to vary the
parameters related to the injection spectrum without appreciable changes
in the final probabilities. Summarizing, HiRes is not unlikely to be
observing events extracted at random from the AGASA spectrum. In other
words, both AGASA and HiRes simply do not have the necessary statistics
of events required to assess either the presence or absence of the GZK
feature in the spectrum of UHECRs.

\section{Combining spectrum and small scale anisotropies}

If UHECRs are accelerated in astrophysical sources, then small scale
anisotropies (SSA) should be observed at the highest energies. The level
of SSA depends on the average density of sources and on the strength of
possible magnetic fields in the intergalactic medium. Simulations of
the formation of the large scale structure of the universe find that
most of the volume of the universe is very weakly magnetized, and the
typical deflections at energies above $4\times 10^{19}$ eV are smaller
than experimental accuracy \cite{dolag04} (however, see \cite{sigl} for
a different conclusion). UHECR Astronomy should  become possible at the
highest energies starting with the detection of either small or large
scale anisotropies. Observations of SSA can be used to infer the average
density of sources and the typical luminosity of each source, an
information which is not accessible through the spectrum of diffuse
UHECRs alone.

\begin{figure}
   \centering
   \includegraphics[width=0.9\textwidth]{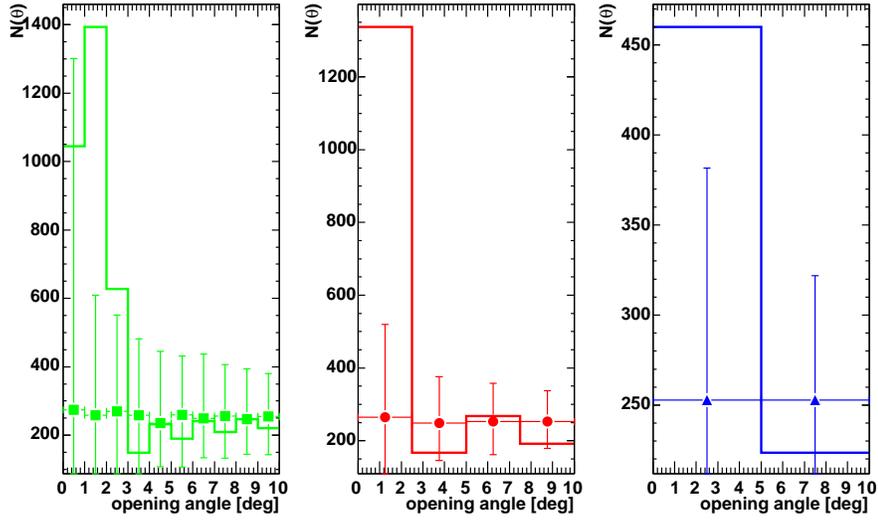}
   \caption{Two point correlation function for AGASA data (histograms)
and Monte Carlo simulations (points with error bars). The latter are
calculated for a homogeneous distribution of sources.}\label{fig:2point}
\end{figure}

AGASA has reported the detection of SSA \cite{agasa_ssa}, whose
significance has been questioned in \cite{finley,cronin}. Even if not
significant, the SSA data from AGASA can be used to exemplify the power
of combining SSA and the spectrum.  In \cite{BD04},  the best fit
density of sources based on AGASA SSA   was estimated to be $\sim
10^{-5}\Mpc^{-3}$, although with large uncertainties (see also
\cite{others}).  In Fig. \ref{fig:2point}, we plot the two point
correlation function, defined as in \cite{BD04} for different choices of
the angular binning over which the multiplets are searched for (1 degree
in the left panel, 2.5 degrees in the central panel, and 5 degrees in
the right panel). The points with error bars are the expected
(simulated) two point correlation function on the given angular scale
for a homogeneous distribution of sources. It is clear that the
statistical significance of the small scale anisotropies decreses
significantly for angular binning on scales smaller or larger than $\sim
2$ degrees.

In Section 3, we concluded that the measurements of the spectrum by
AGASA and HiRes collaborations are not yet statistically significant
around the GZK  feature. In those calculations, we used the assumption
of a uniform  distribution of  diffuse sources. We repeat these
calculation for a discrete distribution of sources with density
$10^{-5}\Mpc^{-3}$. With this source distribution, the average
separation between sources is of the order of $\sim 46$ Mpc, such that
the shape of the GZK feature should be more pronounced than for the
uniform case. As a consequence, the disagreement of AGASA with the
prediction of a GZK suppression should become more pronounced. In order
to demonstrate this point, we simulate 50000 realizations of the AGASA
statistics above $4 \times 10^{19}\eV$, in a scenario of discrete
sources with density $10^{-5}\Mpc^{-3}$. In this case, we find 6
realizations out of the 50000 producing 11 or more events above
$10^{20}\eV$ and this corresponds to a probability of about $10^{-4}$ or
$3.7\sigma$. Clearly, the result is less (more) severe for source
densities $10^{-4}\Mpc^{-3}$ ($10^{-6}\Mpc^{-3}$).

\begin{figure}
   \centering
   \includegraphics[width=0.9\textwidth]{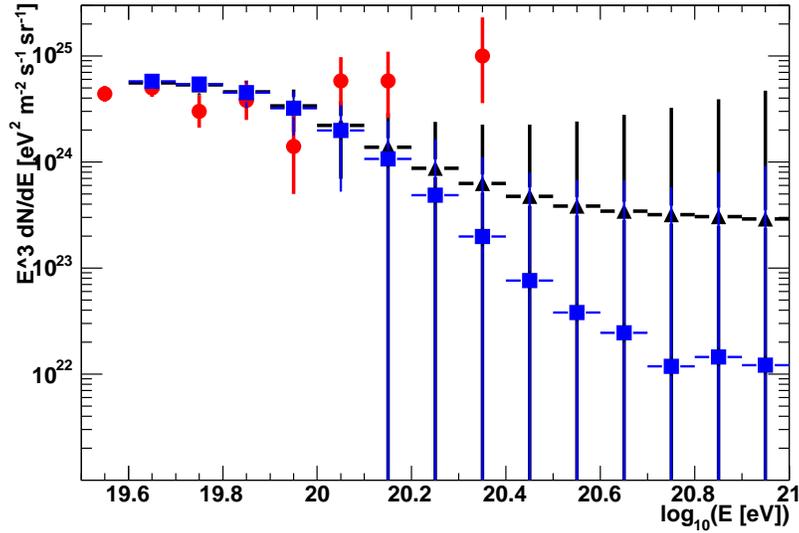}
   \caption{AGASA spectra multiplied by $E^3$. The triangles 
are the result of our simulation in the scenario of continuous 
distribution of sources, the squares are the result of 
simulations with discrete sources with density of $10^{-5}\Mpc^{-3}$ 
and the dots are the AGASA data.
   }\label{fig:agasaspectra}
\end{figure}

This result is further illustrated in Fig. \ref{fig:agasaspectra} where
we show a comparison between the spectra obtained for the AGASA
statistics of events for a homogeneous distribution of sources (black
triangles  with error bars) and point sources with  density
$10^{-5}\Mpc^{-3}$ (blue squares with error bars). The circles are the
AGASA data.

\begin{figure}
   \centering
   \includegraphics[width=0.9\textwidth]{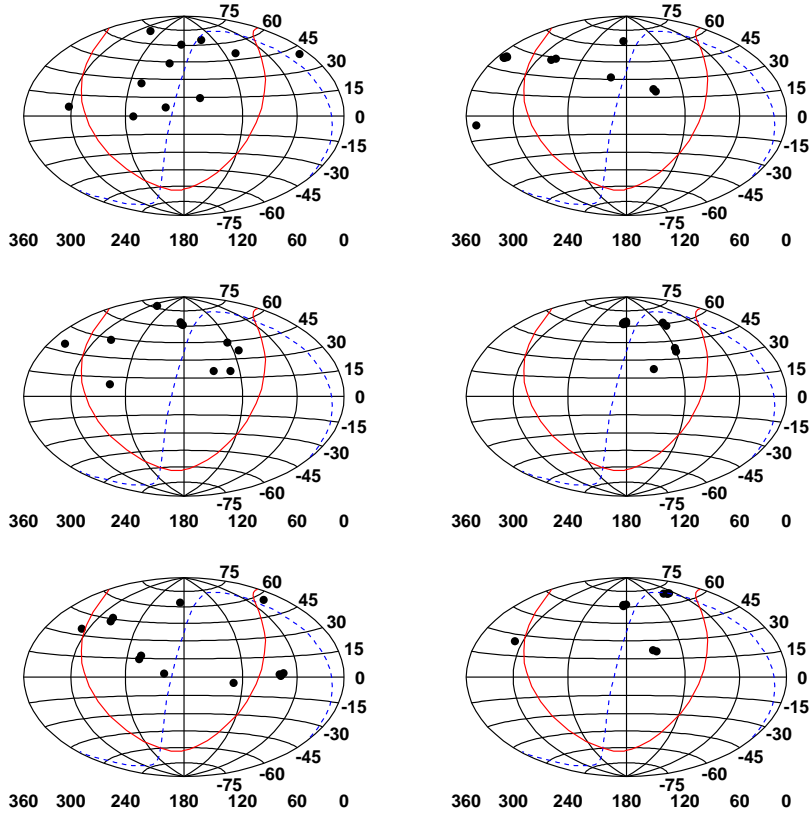}
   \caption{In the above panels we plot the sky distribution of the 
simulated events in the 6 realizations that have 11 or more 
events above $10^{20}\eV$.}
   \label{fig:sky}
\end{figure}

In conclusion, the spectrum of AGASA is inconsistent  (at the $\sim 4
\sigma$ level) with a GZK feature generated by the source distribution
implied by the reported SSA. These discrepancies could be  more severe
if the maximum energy at the accelerators is lower than that assumed
above. In addition,  the distribution in the sky of the 11 (or more)
events in these realizations are not necessarily similar to the AGASA
case. We show our simulated sets of events in Fig. \ref{fig:sky}. With
the exception of the first map in the top left corner, all the
realizations appear to have a higher number of multiplets when compared
with AGASA observations. In other words, if UHECR sources are
astrophysical proton accelerators,  there is an inconsistency between
the spectrum and small scale anisotropies of the AGASA experiment. If we
require that both the spectrum and the sky distributions are correctly
reproduced with a source density $10^{-5}\rm Mpc^{-3}$, the probability
that this occurs in the AGASA data is $\sim 2\times 10^{-5}$. This
exercise shows that the combination of the spetral and anisotropy
information is a powerful tool to estimate the density of sources of
UHECRs and identify possible internal inconsistencies in the data. As
the wealth of data expected from Auger becomes available in the next few
years, this type of study should  help the identification of UHECR
sources.

\section{Conclusions}

We investigated numerically the propagation of UHECRs using a Monte
Carlo code originally developed in \cite{DBO03} and generalized to
include astrophysical point sources in \cite{BD04}. The code  contains
all the main particle physics aspects of the propagation, as well as
several detector-related features (exposures, angle and energy dependent
acceptances, statistical error in the energy determination, and limited
statistics of events). We performed  simulations constrained by the
number of observed events above $4 \times 10^{19}$ eV and compared the
number of events above $ 10^{20}$ eV  observed by AGASA and HiRes. For
AGASA we jointly analyzed the observed spectrum and small scale
anisotropies. The combination of the two types of information
strengthens the constraints on the presence of a the GZK feature or of
clustering of events on small angular scales.

In conclusion, we  first presented the results of our simulations of the
AGASA and HiRes statistics of events similar to the analysis in
\cite{DBO03}. We followed an alternative route and reached very similar
conclusions.  We simulated 30000 realizations with different scenarios
and  showed that the probability distribution of these events is
non-gaussian.  When expressed in  terms of gaussian standard deviations
the probability to record 11 or more events with energies above
$10^{20}$ eV corresponds to $\sim 3\sigma$. If a systematic energy
decrease of 15\% is introduced, the probability of a AGASA spectrum
becomes $\sim 2\%$. We also found that the probability of finding a
spectrum similar to HiRes (one or less than one event at energies above
$10^{20}$ eV) is 3\%. The two spectra cannot therefore be considered
inconsistent with each other at a level better than $2\sigma$.

We also analyzed the small scale anisotropies as detected by AGASA to
stress the fact (already found in \cite{finley}) that the statistical
significance of the clustering depends on the angular scale with which
the data are binned. We clarified this issue by calculating the two
point correlation function for the AGASA data and for the simulated data
as obtained from a  homogeneous distribution of sources. It is clear
from this result that the angular binning chosen by AGASA is the one
that optimizes the small angle clustering signal, but the statistial
significance of such signal decreases for smaller or larger angular
binnings.

Finally, we combined the spectrum and the small scale clustering
information. Choosing a source density of $10^{-5}\rm Mpc^{-3}$ obtained
by fitting the number of doublets and triplets of AGASA (assumed to be
real), we recalculate the spectrum of UHECRs. As expected, for this
source density the GZK feature is more pronounced than that  obtained
for a homogeneous distribution of sources and the discrepancy with
respect to the AGASA spectrum is at the level of $\sim 4\sigma$.
Moreover, by simulating 50000 realizations of the propagation, we
obtained the arrival directions of the events in those realizations that
show a number of events above $10^{20}$ eV equal to or larger  than that
of AGASA (11 events).  Only 6 out the  50000 realizations fulfill the
spectrum requirement  and of the 6 realizations only one has a sky
distribution of the events with resembles the AGASA one (no multiplets).
We conclude that, if the sources of UHECRs are astrophysical proton
accelerators, the reported small scale anisotropies are not compatible
with the spectrum detected by the AGASA experiment at the level of
$2\times 10^{-5}$ probability. In other words only 1 realization out of
50000 has a spectrum and a sky distribution of the events that are
compatible with the AGASA data.

\ack 
The work of D.D.M. is funded through NASA APT grant NNG04GK86G at
University of Delaware. The research of P.B. is funded through
COFIN2004/2005. This work was supported in part by the KICP under NSF
PHY-0114422, by  NSF  AST-0071235, and  DE-FG0291-ER40606 at the
University of Chicago.

\end{document}